# Low-temperature fabrication of brown $TiO_2$ with enhanced photocatalytic activities under visible light

Mingzheng Wang,[1,2] Ka-Kit Yee,[3] Biao Nie,[2] Hua Cheng,[2] Jian Lu,[4,5] Linbao Luo,[1] Zhengtao Xu,[3] Yang Yang Li [2,6,7,*]

[1] School of Electronic Science and Applied Physics, Hefei University of Technology, Hefei, Anhui 230009, China

[2] Center Of Super-Diamond and Advanced Films (COSDAF), [3] Department of Biology and Chemistry, [4] Department of Mechanical and Biomedical Engineering, [5] Center for Advanced Structural Materials, [6] Centre for Functional Photonics, [7] Department of Physics and Materials Science, City University of Hong Kong, Kowloon, Hong Kong SAR, China

* Tel: 852-3442-7810; E-mail: yangli@cityu.edu.hk

**Abstract**

Titanium dioxide is a photocatalytic substance of great practical importance. However, with its bandgap in the ultraviolet (UV) regime, native forms (undoped) of $TiO_2$ generally exhibits poor photocatalytic activities under visible light. Here we report a facile one-step low-temperature method to treat native $TiO_2$ with NaH in a solution-based protocol. The NaH treatment effectively induces the Ti(III) species and oxygen vacancies into the $TiO_2$ host lattice, and enables the bandgap of $TiO_2$ to be conveniently adjusted from the UV region to the red end of the visible spectrum. The modified $TiO_2$ exhibited significantly enhanced photocatalytic capability under visible light, and lead to faster photo-degradation of organic chemical material. Compared with other ways to reduce the bandgap of $TiO_2$, the approach reported here provides unique advantages for safe, large-scale and economic production of narrow-bandgap $TiO_2$ materials.

1. **Introduction**

Titanium oxide ($TiO_2$), as a type of widely studied photocatalyst, offers attractive technological advantages, but presents outstanding limitations at the same time. The former includes high stability, low toxicity and wide commercial availability—advantages that are crucial for large-scale applications in energy and environment areas (e.g., hydrogen production and water decontamination).[1-3] Native forms of $TiO_2$, however, have a wide electronic bandgap (e.g., 3.0 and 3.2 eV for rutile and anatase); as a result, only UV or higher-energy photons can be effectively absorbed by native $TiO_2$, while the visible light and longer-wavelength photons cannot. As UV light only covers a small fraction (3-5%) of the solar spectrum (the visible/infrared region is dominant), $TiO_2$ in the native forms is generally not efficient in harnessing solar energy for photocatalytic applications.

Intensive efforts are therefore being made to reduce the bandgap of $TiO_2$, in order to achieve higher responsiveness to the abundant visible light in the solar spectrum. Several strategies have been reported. One common strategy is to mix in additives or reductants with the titanium compound precursors in the preparative process so that the resultant $TiO_2$ is doped with $Ti^{3+}$ or other impurity centers.[4-11] This can be accomplished using various approaches, e.g., the sol-gel[4,8,12-15] or hydrothermal[16,17] methods. However, methods of this kind often involve expensive chemicals and complex procedures.

A more convenient bandgap engineering approach that is being intensively investigated is to directly modify native TiO$_2$. This is usually done by heating native TiO$_2$ in a particular atmosphere (e.g., H$_2$, NH$_3$ or N$_2$).[18-23] By means of thermal treatments, TiO$_2$ products doped with various impurity atoms (e.g., N),[22] self-doped with Ti$^{3+}$ centers[18] have been successfully made. Among these, H$_2$-thermal treatment[18-21, 24, 25] has lately attracted much attention as a particularly effective method to generate TiO$_2$ with exceptionally narrow bandgap and outstanding photocatalytic properties in solar or visible light. Mechanisms accounting for the bandgap narrowing have been suggested to involve the emergence of the Ti$^{3+}$ ions or oxygen vacancies in the H$_2$-treated TiO$_2$ samples.[18,20,21,24] In addition, a recent breakthrough reported by Chen *et al.*[19,20] pointed to a novel bandgap engineering mechanism: through the introduction of surface disorders.[19,20,24,26] In spite of these encouraging progress achieved, the reported hydrogenation procedures generally involve high safety risk due to the application of high-temperature (e.g., 500 °C) or high-pressure (e.g., 20 bar) H$_2$.

Here we report a facile solution-based method to modify native TiO$_2$ via treatment with a NaH solution under the mild condition of low temperature (e.g., 150 °C) and atmospheric pressure, thus obviating the safety hazard associated with the reported hydrogenation methods that entail high-temperature/pressure H$_2$. Using this solution protocol, doped TiO$_2$ solid samples featuring adjustable bandgaps (from UV to the red) and largely enhanced photo-activities in visible light, can be conveniently prepared. In particular, brown-colored (Fig. 1) TiO$_2$ solids can be fabricated, adding a new hue to the color selection[7,19,20,24,27] of bandgap-engineered TiO$_2$ products.

## 2. Experimental

*2.1 Fabrication of brown TiO$_2$*     Brown TiO$_2$ was prepared by a one-step solution-based method under N$_2$ protection. Inside a glove box, the mineral oil-NaH mixture (Acros, 65 wt.% NaH) was rinsed with anhydrous hexane to obtain NaH powder. 10 ml of anhydrous dimethylformamide (Aldrich, DMF) was added to a quartzose round-bottom flask containing 100 mg of TiO$_2$ (Degussa, P-25) and 200 mg of NaH powder. The flask was then taken out of the glove box and connected to a Schlenk line. The suspension in the flask was stirred and heated at an elevated temperature (e.g., 150 $^o$C) in N$_2$ for several hours (e.g., 4 hrs). To investigate the impact of UV irradiation on the reaction rate, some samples were prepared with a mercury lamp (maximum output wavelength at 365 nm) placed at a distance of 20 cm from the flask during the reaction. After reaction, the flask was cooled down to room temperature. The insoluble material was collected by centrifugation, and then sequentially washed with 2-propanol (International Laboratory USA, 99.7%), anhydrous acetone (Sigma-Aldrich, 99.99%) and absolute ethanol (Sigma-Aldrich) each for three times and finally dried in a nitrogen stream. The representative samples and their fabrication parameters are shown in Table 1.

Table 1. Reaction parameters and resultant bandgaps for NaH-treated TiO$_2$ samples.

| Sample index | Reaction temperature | Reaction time | UV illumination for reaction | Bandgap measured |
|---|---|---|---|---|
| P-25 (native) | — | — | — | 3.04 eV |
| Sample A | 150 °C | 4 hrs | off | 2.69 eV |
| Sample B | 120 °C | 4 hrs | on | 2.32 eV |
| Sample C | 150 °C | 2 hrs | on | 2.11 eV |
| Sample D | 150 °C | 4 hrs | on | 1.82 eV |

*2.2 Characterizations*   X-ray diffraction (XRD) measurements were performed using an X-ray diffractometer (Philips X'pert). Morphology studies were carried out using a transmission electron microscope (TEM) (Philips CM20, operating at 200 kV). Chemical compositional analysis was carried out using an energy dispersive X-ray spectrometer (EDS) equipped in the TEM. Surface chemical analysis was measured using an X-ray photoelectron spectrometer (XPS) (VG ESCALAB 220i-XL) equipped with a monochromatic Al Kα (1486.6 eV) source. Optical properties were investigated using a diffuse reflectance UV-Vis absorption spectrophotometer (Perkin Elemer Lambda λ750) equipped with an integrating sphere attachment. The spectra were recorded at room temperature from 200 to 800 nm. The BaSO$_4$ standard mirror was used as the reference.

*2.3 Photo-activity measurements*   The photo-degradation abilities of the TiO$_2$

samples were tested as follows. Typically, 20 mg of the $TiO_2$ sample was added into a 20 ml aqueous solution of Phenol (20 mg/L). The mixture was stirred for 30 min in dark to allow the establishment of the absorption-desorption equilibrium between the Phenol and $TiO_2$. For the UV illumination source, a mercury lamp with the maximum output wavelength of 365 nm was used. For the visible light source, a 300 W tungsten lamp equipped with a UV cutoff ($\lambda > 420$nm) filter was used. The lamps were kept at a distance of 15 cm from the reaction vessel with the light intensity there measured to be 206 mW/cm$^2$. The diameters of reaction vessels were 4.0 cm. When the visible light source was used, a water filter was placed between the lamp and the reaction vessels for absorbing the heat and the infrared radiation over ~ 1100 nm. The temperature was maintained around 25 °C during the photocatalytic reaction. The Phenol concentration in the reaction solution ($TiO_2$ removed by centrifugation beforehand) was measured at ~ 270 nm by the UV-Vis spectrometer (Shimadzu UV-1700).

## 3. Results and Discussion

Distinct color change from white to brown was readily observed in the above treatment of $TiO_2$ by NaH (i.e., Sample D; see Fig. 1). The TEM study (Fig. S1a-c) revealed that the particle sizes of native (P-25) and treated (sample D) $TiO_2$ particles were in the range of 20 ~ 50 nm, whereas the elemental analysis of both samples by EDS (Fig. S1d) suggested a chemical formula consistent with $TiO_2$ (i.e., the degree of doping was not detectable by EDS).

XRD patterns (Fig. 2) of the commercial native $TiO_2$ (P-25) and NaH-treated $TiO_2$ (Samples A-D) indicated that the crystallinity of the $TiO_2$ solid was maintained before and after the treatment, with no impurity peaks observed. However, the diffraction peaks of Samples A-D shifted to the larger diffraction angles comparing to P-25 (the shift was particularly more apparent for the higher-order peaks), indicating a smaller lattice spacing in the NaH-treated samples. Furthermore, lower peak intensities were observed on the treated $TiO_2$ (particularly Sample D), suggesting an increased level of defects resulted from the treatment. This observation is in good agreement with the $TiO_2$ samples doped using other methods.[19]

To investigate the chemical environment and states of Ti and O in the treated $TiO_2$, XPS measurements were performed. The XPS spectra of native $TiO_2$(P-25) and treated (Sample D) $TiO_2$ indicated that only Ti, O elements and a trace amount of carbon were present in the samples. The high-resolution Ti2p spectra (Fig. 3a) indicated an apparent shift towards the lower energy end for the peaks of NaH-treated $TiO_2$ (457.2 eV), relative to native $TiO_2$ (458.6 eV). This observation is thus in line with the presence of $Ti^{3+}$ in the NaH-treated $TiO_2$ samples, for the major Ti2p peak for $Ti^{3+}$ is normally found to at lower energy (around 457.7 eV) than that of $Ti^{4+}$ (459.5 eV).[18] Moreover, the XPS study also revealed the formation of oxygen vacancies resulted from the NaH treatment. The O1s spectra (Fig. 3b) distinctly showed a new peak emerging at 531.2 eV, which can be ascribed to the existence of oxygen-vacancy sites nearby $Ti^{3+}$.[7, 26]

The UV-Vis diffuse reflectance spectra (Fig. 4a) showed that, comparing to native $TiO_2$, the absorption edges of NaH-treated $TiO_2$ shifted to longer wavelengths in the visible region. The spectrum of native $TiO_2$ showed a single steep absorption edge around 400 nm corresponding to its bandgap. Clearly, NaH-treated $TiO_2$ exhibited different optical responses. In particular, Sample D displayed considerable absorption in the visible region between 400 and 700 nm, consistent with its brown-colored appearance.

The Kubelka-Munk function, $F(R)$, as a function wavelength was derived from the measured diffuse reflectance spectra, based on the Kubelka-Munk equation[28]:

$$F(R) = \frac{(1-R)^2}{2R}$$

where $R$ is the ratio of the sample reflectance to the reference reflectance. A Tauc plot ($[F(R) \bullet h\nu]^{1/2}$ vs. $h\nu$, where $h\nu$ is the photon energy) was then obtained (Fig. 4b). The bandgap ($E_g$) was determined from the extrapolation of the linear fit for the Tauc plot onto the energy axis.[29] It was estimated that $E_g$ is 3.04 eV for native $TiO_2$ and 2.69, 2.32, 2.11 and 1.82 eV for Samples A-D, respectively (Table 1). Previous studies show that, as the concentration of the $Ti^{3+}$ ions/oxygen vacancies increase, the band tailing becomes more evident and the narrower bandgap is resulted.[18,24] The gradual narrowing in bandgap from Samples A to D indicates the gradual increasing concentration of the $Ti^{3+}$ ions/oxygen vacancies in these samples. The finding that P-25 (native $TiO_2$) showed a smaller bandgap than the bulk anatase ($Eg$ = 3.2 eV) is in good agreement with previous studies, and can be ascribed to the fact that that P-25 is mix-phased nanoparticles containing ~ 25% rutile phase and 75% anatase phase.[8]

A possible reaction mechanism leading to the formation of the $Ti^{3+}$ ions and the oxygen vacancies by the NaH treatment are shown in eqs. 1-3. First, the $H^-$ ions supplied by NaH bond with $Ti^{4+}$ to form a titanium hydride species $[TiH]^{3+}$; the $[TiH]^{3+}$ species then disintegrate to give the $Ti^{3+}$ centers and H· radicals that readily combine to form $H_2$ gas. The charge balance is likely maintained by intercalation of $Na^+$ ions into the host lattice of $TiO_2$, generating confined defects featuring local stoichiometry of $NaTiO_2$.

$$Ti^{4+} + H^- \rightarrow [Ti-H]^{3+} \qquad (1)$$

$$[Ti-H]^{3+} \rightarrow Ti^{3+} + \bullet H \qquad (2)$$

$$2 \bullet H \rightarrow H_2 \uparrow \qquad (3)$$

A comparison between Samples A and D (0.9 eV difference in bandgap width) indicates that the UV irradiation applied during the NaH treatment can result in a markedly smaller bandgap in the treated $TiO_2$ product. Previous studies has shown that trace amount of the Ti(III) species can be produced transiently in irradiated $TiO_2$.[30,31] The significant impact of UV irradiation observed here could be attributed to the UV-photon generated electrons and holes in $TiO_2$. The photogenerated electron-hole pairs may facilitate the doping process (see eqs. 4-6), e.g., the positive holes generated in the valence band of $TiO_2$ is highly oxidizing, and therefore more readily accepts electrons from the $H^-$ ions compared with the $TiO_2$ system without excitation by UV light. Furthermore, because the electron-hole pairs can be UV-excited not only at the surface layer but also inside the $TiO_2$ lattice, $Ti^{3+}$ can be produced (eq. 5) deeper inside the $TiO_2$ lattice. Comparing to those on the surface, the $Ti^{3+}$ ions inside the $TiO_2$ are likely much less susceptible to oxidation by oxygen,[27] leading to a higher stability of the treated $TiO_2$ when placed in the ambient environment.

$$TiO_2 + h\nu \rightarrow TiO_2 + e^-_{CB} + h^+_{VB} \quad (4)$$

$$Ti^{4+} + e^-_{CB} \rightarrow Ti^{3+} \quad (5)$$

$$h^+_{VB} + H^- \rightarrow \cdot H \quad (6)$$

Comparison between Samples B and D shows that a higher reaction temperature (e.g., by 30 °C) can lead to a substantial reduction of bandgap (e.g., by ~0.5 eV). In general, the reductive doping of the very stable $TiO_2$ solid is a slow process. The great effects of higher temperature and UV treatment in reducing the bandgaps of the doped $TiO_2$ product help to highlight the underlying reasons for the success of the current strategy for self-doping $TiO_2$.

Compared with native $TiO_2$, the NaH-treated $TiO_2$ samples all displayed higher ability to photo-degrade Phenol, with the degradation rates gradually increasing from Sample A to Sample D (Fig. 5). To explain the higher photo-degradation rate observed on NaH-treated $TiO_2$, a discussion on the photo-degradation mechanism is helpful. $TiO_2$ is an n-type semiconductor. For native $TiO_2$, when illuminated with UV photons with energy higher than its bandgap, the photo-excited electrons in the conduction band (CB) can be transferred to the $O_2$ molecules absorbed on the surface of $TiO_2$ and form oxygen radicals, $•O_2^-$, that can degrade the absorbed organic molecules. For treated $TiO_2$, previous studies indicate that the $Ti^{3+}$ ions/oxygen vacancies induce donor levels in the bandgap (a higher concentration of the $Ti^{3+}$ ions/oxygen vacancies leads to more donor levels in the band gap).[18,24] These donor levels may help trap the photogenerated electrons which generate $•O_2^-$, lessening the recombination of the photogenerated electrons and holes. Therefore, under UV irradiation, the existence of the donor states bestows upon NaH-treated $TiO_2$ enhanced photocatalytic activity, with the wider tail band leading to the greater enhancement.

Notably, the enhancement effect on photodegration under UV irradiation is not as dramatic as under visible illumination (Fig. 5). This is because the photon energies of visible light are insufficient to excite native $TiO_2$ but high enough to excite NaH-treated $TiO_2$. It was found that $TiO_2$ with a lower bandgap displayed higher photodegrading efficiency, possibly because they utilized a larger portion of the visible spectrum and possessed more donor levels to trap more photo-excited electrons which in turn generate more $•O_2^-$.

A closer look at Fig. 5b reveals that, under visible irradiation, the degradation rate for native $TiO_2$ is very low but slightly higher than when no catalyst was used. It is well known that native $TiO_2$ cannot be excited by visible light because of its wide bandgap. This observed slight degradation ability of native $TiO_2$ under the visible light irradiation is possibly caused by the photo-sensitizing capability of the dye molecules which can be excited by the visible light and inject electrons to the conduction band of $TiO_2$.[2]

## 4. Conclusions

In summary, brown-color $TiO_2$ with the bandgap in red light has been conveniently fabricated using a facile one-step treatment in a NaH solution. The dramatic bandgap narrowing is attributed to the $Ti^{3+}$ ions/oxygen vacancies created by the NaH treatment. The fabricated brown $TiO_2$ showed significantly higher photocatalytic activity than untreated commercial $TiO_2$. The outstanding photocatalytic performance and the great fabrication convenience enabled by the novel method presented in this study open a new route to practical applications of $TiO_2$ under visible light.

**Acknowledgments**

This work was supported by the City University of Hong Kong (Projects 9667070 and 7003039) and the National Natural Science Foundation of China (Project 51202206).

**Figure Captions**

Fig. 1 Photographs of commercial (P-25) and NaH-treated (Sample D) $TiO_2$.

Fig. 2 XRD patterns of commercial $TiO_2$ (P-25) and Samples A-D whose fabrication conditions were shown in Table 1.

Fig. 3 High-resolution Ti2p (a) and O1s (b) XPS spectra of commercial $TiO_2$ (P-25) and Sample D.

Fig. 4 Reflectivity spectra (a) and Tauc plots (b) of commercial $TiO_2$ (P-25) and Samples A-D.

Fig. 5 Photodegradation rate of Phenol under UV (a) and visible (b) light., measured with commercial $TiO_2$ (P-25) and Samples A-D.

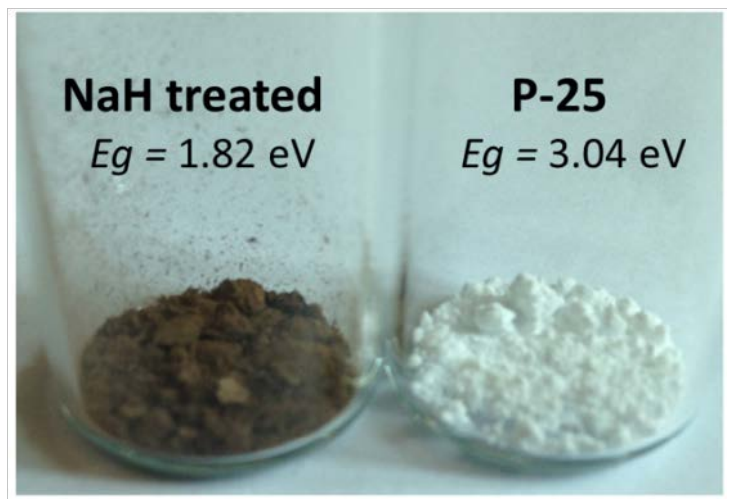

Fig. 1 Photographs of commercial (P-25) and NaH-treated (Sample D) $TiO_2$.

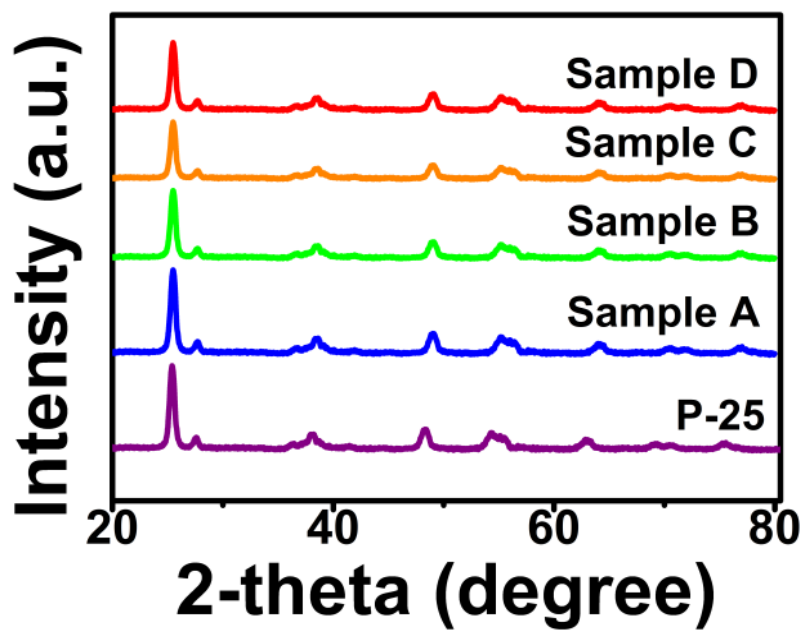

**Fig. 2** XRD patterns of commercial $TiO_2$ (P-25) and Samples A-D whose fabrication conditions were shown in Table 1.

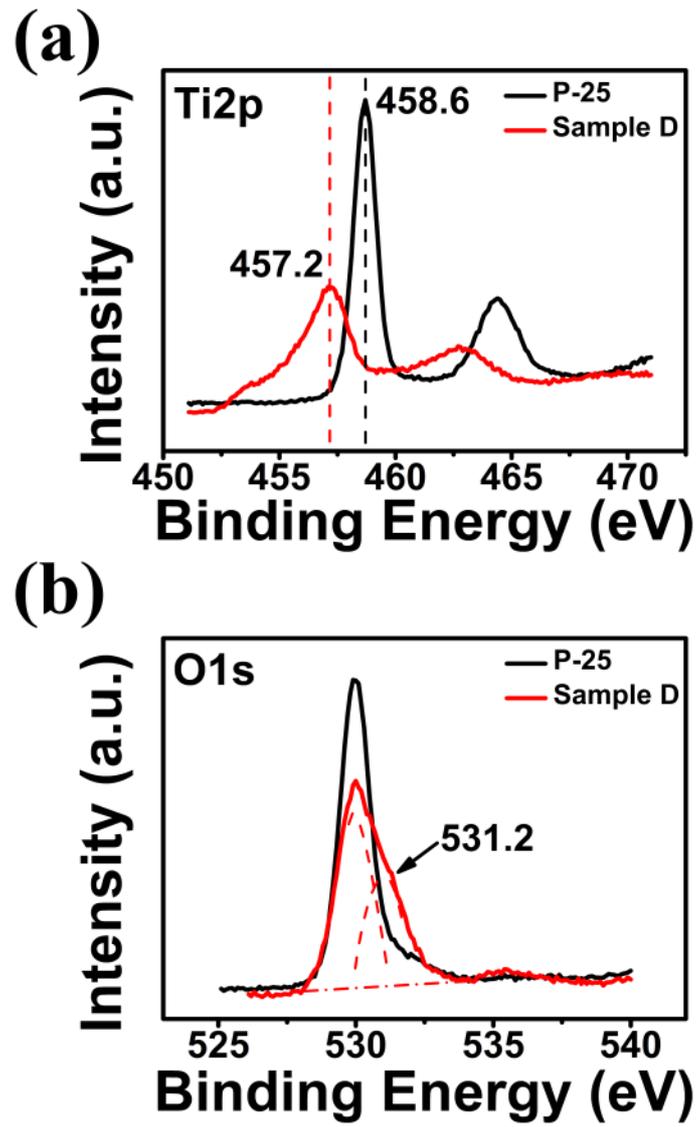

**Fig. 3** High-resolution Ti2p (a) and O1s (b) XPS spectra of commercial TiO$_2$ (P-25) and Sample D.

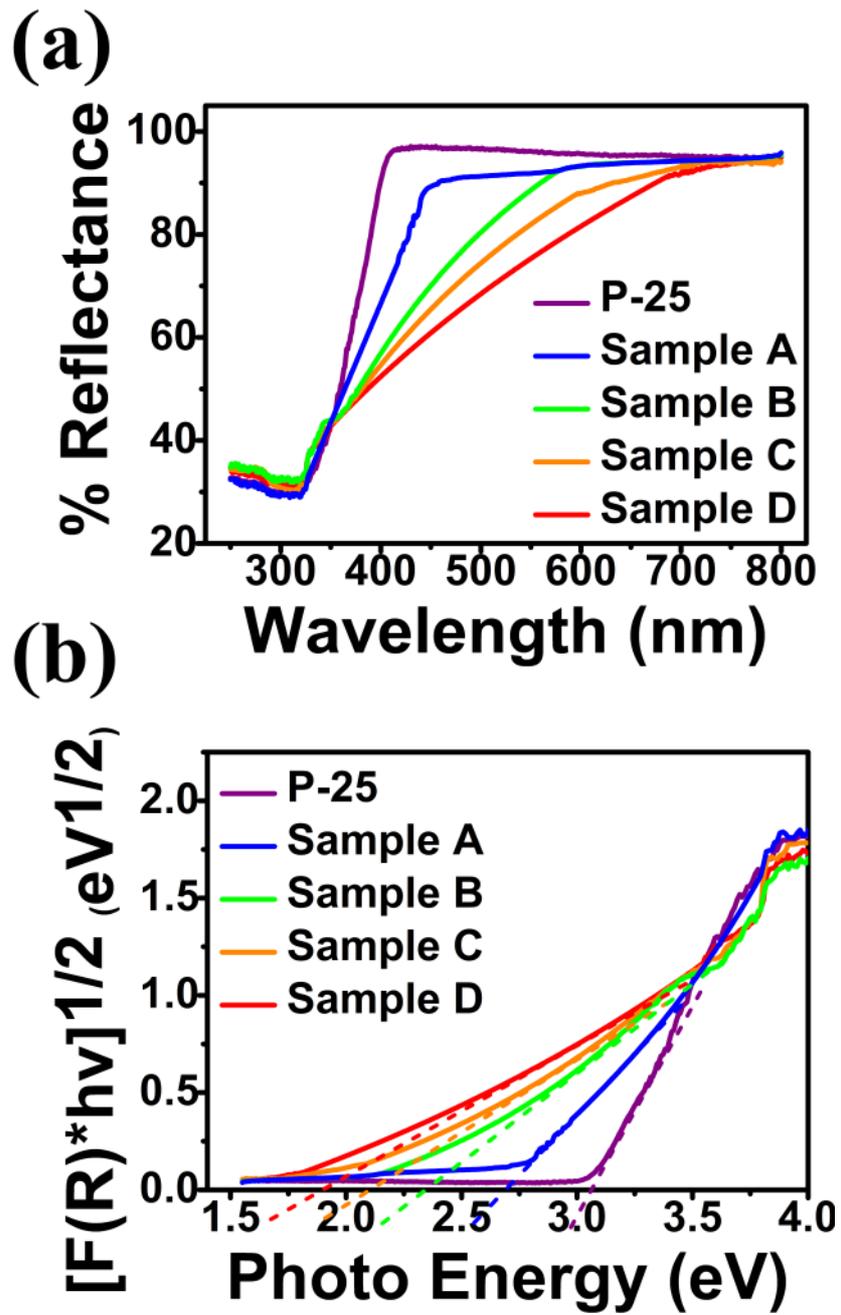

**Fig. 4** Reflectivity spectra (a) and Tauc plots (b) of commercial $TiO_2$ (P-25) and Samples A-D.

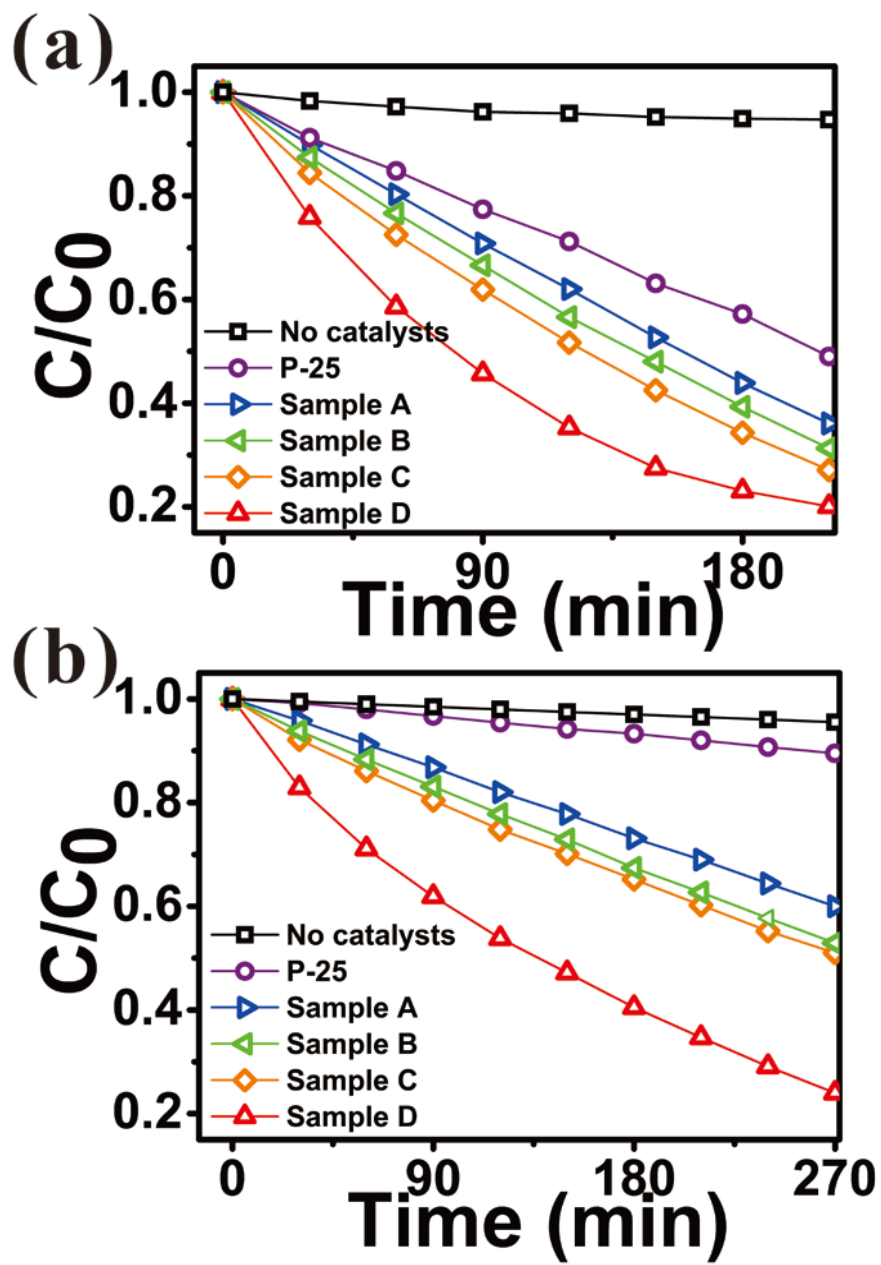

**Fig. 5** Photodegradation rate of Phenol under UV (a) and visible (b) light, measured with commercial TiO$_2$ (P-25) and Samples A-D.

# References

1   Fujishima A., *Nature* **1972**, 238, 37-38.

2   Chen, X.; Mao, S. S., *Chem. Rev.* **2007**, 107, 2891-2959.

3   Carp, O.; Huisman, C. L.; Prog, A. Reller., *Solid State Chem.* **2004**, 32, 33-177.

4   Yu, J. C., *Chem. Mater.* **2002**, 14, 3808-3816.

5   Zuo, F.; Wang, L.; Wu, T.; Zhang, Z.; Borchardt, D.; Feng, P. J., *Am. Chem. Soc.* **2010**, 132, 11856-11857.

6   Ohno, T.; Akiyoshi, M.; Umebayashi, T.; Asai, K.; Mitsui, T.; Matsumura, M., *Appl. Cata. A* **2004**, 265, 115-121.

7   Zheng, Z.; Huang, B.; Meng, X.; Wang, J.; Wang, S.; Lou, Z.; Wang, Z.; Qin, X.; Zhang, X.; Dai, Y., *Chem. Comm.* **2013**, 49, 868-870.

8   Choi, W.; Termin, A.; Hoffmann, M. R., *J. Phys. Chem.* **1994**, 9, 13669-13679.

9   Klosek, S.; Raftery, D., *J. Phys. Chem. B* **2001**, 105, 2815-2819.

10  Chen, X.; Burda, C., *J. Am. Chem. Soc.* **2008**, 130, 5018-5019.

11  Myung, S. T.; Kikuchi, M.; Yoon, C. S.; Yashiro, H.; Kim, S. J.; Sun, Y. K.; Scrosati, B., *Energy Environ. Sci.* **2013**, 6, 2609-2614.

12  Burda, C.; Lou, Y.; Chen, X.; Samia, A. C. S.; Stout, J.; Gole, J. L., *Nano Lett.* **2003**, 3, 1049-1051.

13  Xiao, Q.; Zhang, J.; Xiao, C.; Si, Z.; Tan, X., *Sol. Energy*, **2008**, 82, 706-713.

14  Wang, E.; Yang, W.; Cao, Y., *J. Phys. Chem. C* **2009**, 113, 20912-20917.

15  Kuvarega, A. T.; Krause, R. W. M.; Mamba, B. B., *J. Phys. Chem. C* **2011**, 115, 22110-20120.

16  Yang, H. G.; Liu, G.; Qiao, S. Z.; Sun, C. H.; Jin, Y. G.; Smith, S. C.; Zou, J.; Cheng, H. M.; Lu, G. Q. M., *J. Am. Chem. Soc.* **2009**, 131, 4078-4083.

17  Liu, G.; Yang, H. G.; Wang, X.; Cheng, L.; Pan, J.; Lu, G. Q. M.; Cheng, H. M., *J. Am. Chem. Soc.* **2009**, 131, 12868-12869.

18  Lu, X.; Wang, G.; Zhai, T.; Yu, M.; Gan, J.; Tong, Y.; Li, Y., *Nano Lett.* **2012**, 12 1690-1696.

19  Chen, X.; Liu, L.; Yu, P. Y.; Mao, S. S., *Science* **2011**, 331, 746-750.

20  Chen, X.; Liu, L.; Liu, Z.; Marcus, M. A.; Wang, W. C.; Oyler, N. A.; Grass, M. E.; Mao, B. H.; Glans, P. A.; Yu, P. Y.; Guo, J. H.; Mao, S. S., *Scientific Reports* **2013**, 3, 1510(1)-1510(7).

21  Wang, G.; Ling, Y.; Li, Y., *Nanoscale* **2012**, 4, 6682-6691.

22  Irie, H.; Watanabe, Y.; Hashimoto, K., *J. Phys. Chem. B* **2003**, 107, 5483-5486.

23  Asahi, R.; Morikawa, T.; Ohwaki, T.; Aoki, K.; Taga, Y., *Science* **2001**, 293 269-271.

24  Wang, G.; Wang, H.; Ling, Y.; Tang, Y.; Yang, X.; Fitzmorris, R. C.; Wang, C. C.; Zhang, J. Z.; Li, Y., *Nano Lett.* **2011**, 11, 3026-3033.

25  Naldoni, A.; Allieta, M.; Santangelo, S.; Marelli, M.; Fabbri, F.; Cappelli, S.; Bianchi, C. L.; Psaro, R.; Santo, V. Dal., *J. Am. Chem. Soc.* **2012, 134**, 7600-7603.

26  Leshuk, T.; Parviz, R.; Everett, P.; Krishnakumar, H.; a Varin, R.; Gu, F., *ACS Appl. Mater. & Interfaces* **2013**, 5, 1892-1895.

27  Gordon, T. R.; Cargnello, M.; Paik, T.; Mangolini, F.; Weber, R. T.; Fornasiero, P.; Murray, C. B., *J. Am. Chem. Soc.* **2012,** 134, 6751-6761.

28  Tauc, J., *Mater. Res. Bull.* **1970**, 5, 721-730.

29  Madhusudan Reddy, K., V Manorama, S.; Ramachandra Reddy, A., *Mater. Chem. and Phys.* **2003**, 78, 239-245.

30  Henglein, A.; *Chem. Rev.* **1989**, 89, 1861-1873.

31  Rothenberger, G.; Moser, J.; Graetzel, M.; Serpone, N.; Sharma, D. K.; *J. Am Chem. Soc.* **1985**, 107, 8054-8059.